\documentclass[aps,pre,superscriptaddress,showpacs,twocolumn,floatfix]{revtex4}

\usepackage{graphicx}
\usepackage{url}
\usepackage{color} 
\usepackage{amsmath}
\usepackage{enumerate}
\begin{document}


\title{Unwinding Dynamics of a Helically Wrapped Polymer}


\author{J.-C. Walter}
\affiliation{Laboratoire Charles Coulomb UMR 5221, Universit\'e Montpellier 2 \& CNRS,  F-34095, Montpellier, France}
\affiliation{Instituut-Lorentz, Universiteit Leiden, P.O. Box 9506, 2300 RA Leiden, The Netherlands}
\affiliation{Institute for Theoretical Physics, KU Leuven, Celestijnenlaan 200D, Leuven, Belgium}
\author{M. Baiesi}
\affiliation{Dipartimento di Fisica e Astronomia, Universit\`a di Padova, Via Marzolo 8, Padova, Italy}
\affiliation{INFN, Sezione di Padova, Via Marzolo 8, Padova, Italy}
\author{E. Carlon}
\affiliation{Institute for Theoretical Physics, KULeuven, Celestijnenlaan 200D, Leuven, Belgium}
\author{H. Schiessel}
\affiliation{Instituut-Lorentz, Universiteit Leiden, P.O. Box 9506, 2300 RA Leiden, The Netherlands}



\date{\today}

\begin{abstract}
We study the rotational dynamics of a flexible polymer initially wrapped
around a rigid rod and unwinding from it. This dynamics is of interest in several
problems in biology and constitutes a fundamental instance of polymer relaxation from
a state of minimal entropy. We investigate the dynamics of several quantities such as the
total and local winding angles and metric quantities. The results of simulations
performed in two and three dimensions, with and without self-avoidance, are explained
by a theory based on scaling arguments and on a balance between frictional and entropic
forces. The early stage of the dynamics is particularly rich, being characterized by three
coexisting phases.
\end{abstract}

\maketitle

\section{Introduction}

The genetic information in eukaryotic cells (including cells of animals
and plants) is accessed through DNA unwinding on two different length
scales. On the larger scale the DNA double helix has to unwind from
proteins, on the smaller scale the two strands of the double helix need
to be separated. In the first case a semiflexible polymer (DNA double
helix) is wound almost two turns around a protein cylinder forming
the so-called nucleosome \cite{luger97,schiessel14}, in the second
case two flexible polymers (chains of nucleotides) are twisted around
each other leading to the much stiffer double helix. The unwinding
of the DNA from the nucleosome or the separation of the DNA double
helix is achieved inside a cell in various ways, often involving
molecular motors (chromatin remodellers, polymerases...) that usually
give access through a local opening of the structures. Inside
a test tube unwinding can be induced, typically on a global
scale, through a change in temperature (DNA melting/helix-coil
transition \cite{schiessel14,poland66,kafri02,blossey03}), salt
concentration (salt-induced DNA release \cite{yager87,kunze00})
or through application of an external force (nucleosome unwrapping
\cite{schiessel14,brower-toland02,kulic04,mihardja06,blossey11},
DNA unzipping \cite{kafri02,essevaz97}). Local unwinding can also
occur spontaneously leading to the breathing of nucleosomes
and of DNA (usually called site exposure in nucleosomes
\cite{schiessel14,blossey11,polach95,koopmans09,prinsen10} and
denaturation bubbles in DNA \cite{schiessel14,blossey03,ambj07}). Finally,
during transcription, where the elongating RNA polymerase produces a
RNA transcript, one faces again the situation of a flexible chain, the
transcript, being initially wound around the much stiffer DNA double
helix \cite{belo14}.

To gain insights into the unwinding process it is convenient to start
from simple setups.  In this work we study the unwinding process of
an idealized flexible polymer model that is initially wound around a
stiff rod, in a configuration of minimal entropy, which resembles some
of the features of the DNA or RNA unwinding. This model was introduced
and studied in ref.~\cite{wal13}. Here we extend the results of that
analysis focusing in particular on metric properties. We present a scaling
argument which fully captures the early stages of the dynamics that is
characterized by power-law scaling. Despite the simplicity of the model,
there is a complex dynamical behavior.

\begin{figure}[!htb]
\begin{tabular}{c}
\includegraphics[angle=0,width=0.45\textwidth]{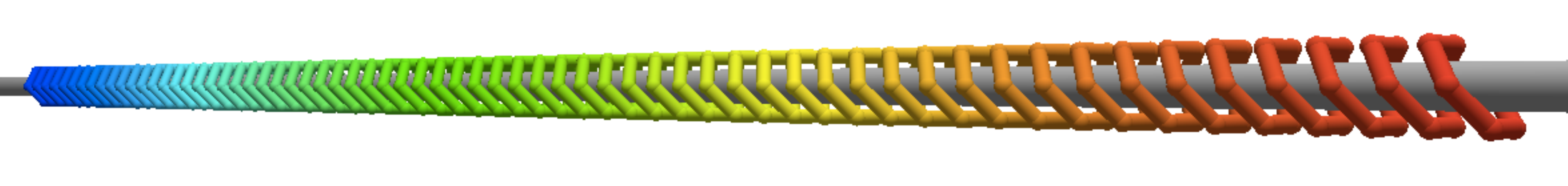}\\
\includegraphics[angle=0,width=0.45\textwidth]{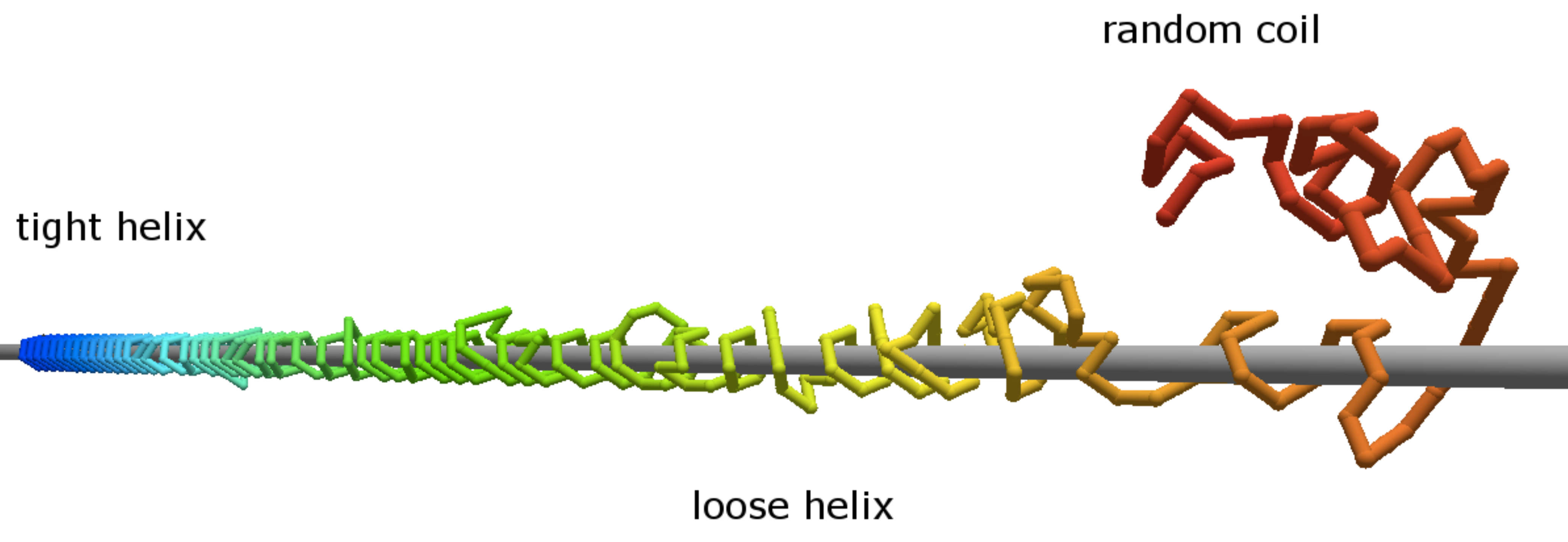}\\
\includegraphics[angle=0,width=0.45\textwidth]{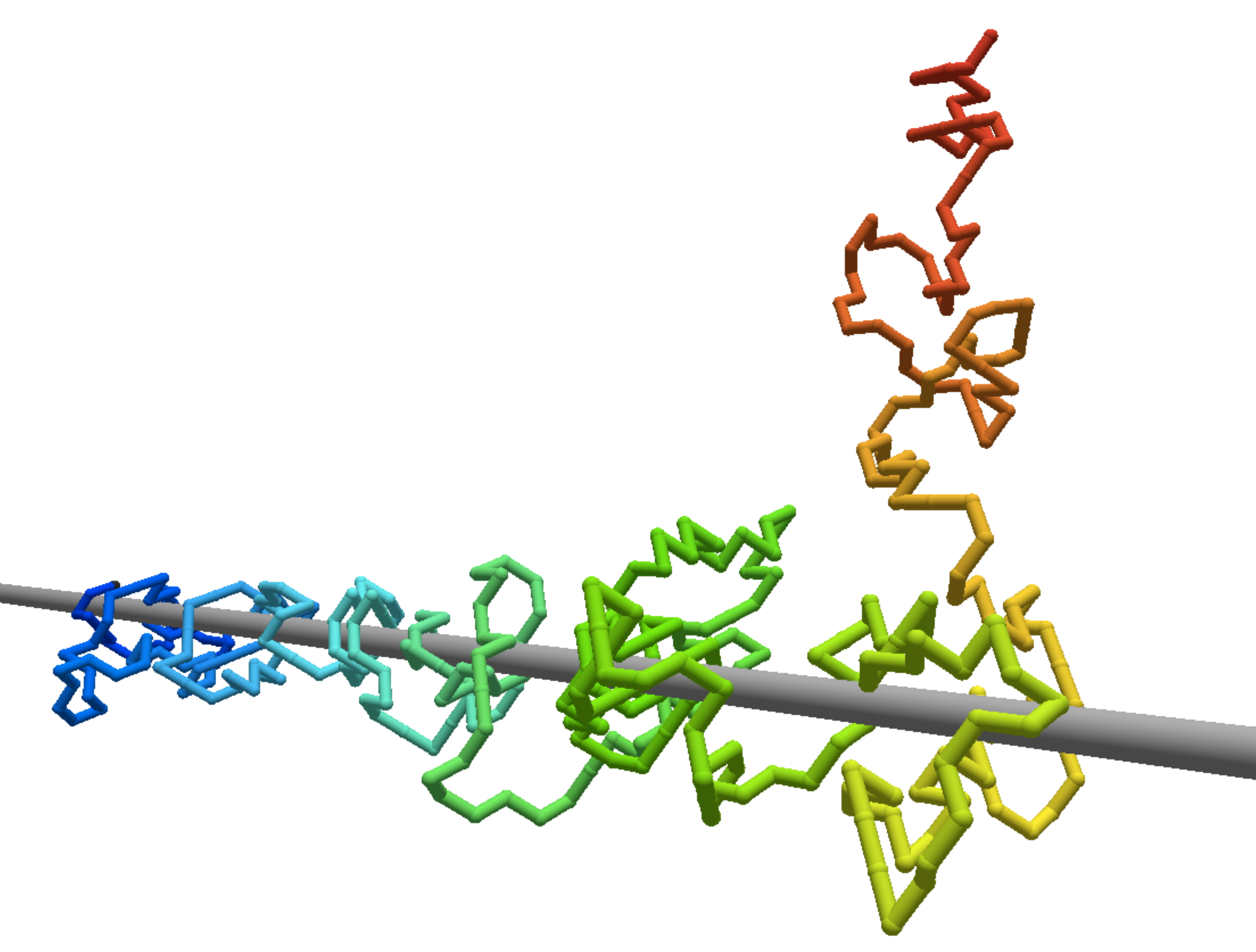}\\
\includegraphics[angle=0,width=0.45\textwidth]{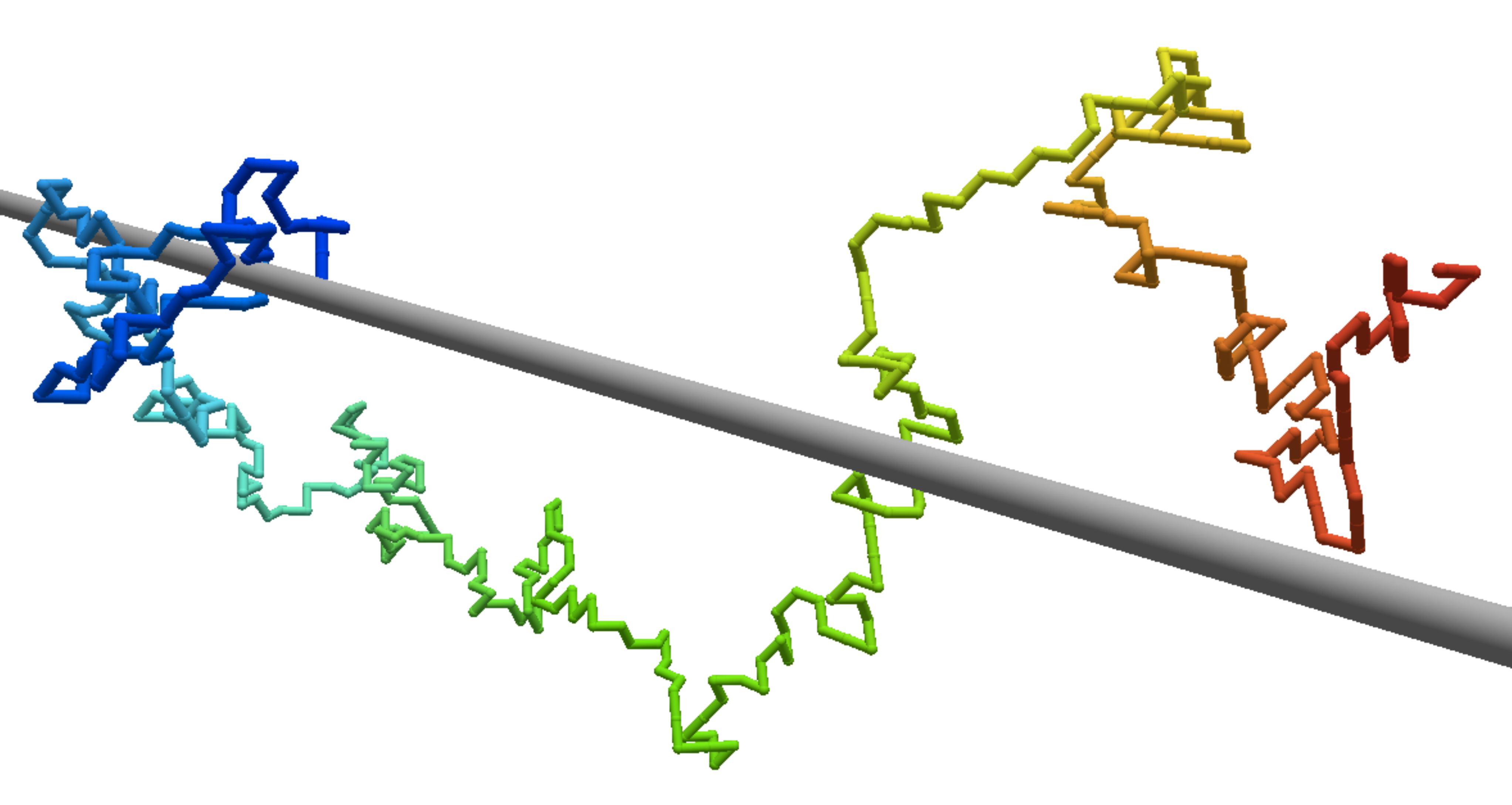}
\end{tabular}
\caption{Snapshots of a SAW with $L=384$ monomers during
unwinding from a rod.  One end (blue) is attached to the rod and
the other end (red) is free. Snapshots are taken at 
times $t=0$, $t=7\times 10^4$, $t=2\times 10^6$, and $t=5\times 10^6$
(from top to bottom). 
The second configuration is in the early stage of the unwinding, where one can
recognize three different phases: the part close to the
fixed end is still a tightly wound helix,
the middle section shows a loose helix configuration,
and the part close to the free end forms a random coil.  }
\label{fig:1}
\end{figure}

It is the entropy gain that drives the polymer from the initial
configuration toward the full random coil configuration. A sketch
of this process is given in Figure~\ref{fig:1}, which shows different
snapshots of the polymer configurations in the course of a simulation.
The polymer is initially fully wound in a helix around the rigid rod.
One end of the polymer is tethered to the rod whereas the other end is
free, hence the relaxation proceeds from the free end.  This process
shares some similarities with the simpler problem of the relaxation from
one end of a completely stretched polymer \cite{brochard95,buguin96}.
Consider a polymer tethered at one end and fully stretched by a strong
flow. When the flow is turned off the polymer relaxes back to its coiled
equilibrium conformation. As one end is tethered the relaxation starts
from a free end, from where the coil grows. The unwinding has a similar
relaxation from the free end which occurs through a rotational motion
instead of a translational recoiling. However, the phenomenology in the
case of unwinding is much richer, as we will show.

This paper is organized as follows. In the section Models and Simulations,
we review the models and the type of Monte Carlo simulations used. In the
section Results and Discussion, we focus first on the early unwinding
dynamics which is characterized by power-laws scaling in time. It is
shown that force balance equations and scaling arguments yield exponents
in very good agreement with simulations.  Second, we discuss the late
stage of the relaxation process. Here the theory of ref.~\cite{wal13}
predicts a power-law scaling for the longest relaxation time with
logarithmic corrections.  The analysis is extended to other lattice and
off-lattice models and the results confirm the validity of the theory.

\section{Models and Simulations}
\label{sec:sim}

Figure~\ref{fig:1} shows different snapshots of the polymer configurations
during the unwinding from an infinitely long rod. Initially the polymer
is fully wound and in the course of time it unwraps. The polymer has $L$
monomers labeled with indices $1 \leq k \leq L$.  One end ($k=1$) is
fixed to the rod, while the other end ($k=L$) is free.  We have studied
different cases to check the robustness and universality
of the numerical results. For the
ideal chain (the case without excluded volume) we modeled the polymer
in three different ways.  In a first model we considered a random walk
(RW) with unit steps on a face-centered cubic (FCC) lattice.  In the
initial helical configuration, the polymer performs one turn in six
steps along the rod.  An update of the configuration consists of $L$
attempts of so-called corner flips, where the randomly selected monomer is
moved to a neighboring lattice site if the distances to the neighboring
monomers are conserved (this is a lattice realization of Rouse dynamics~\cite{doi86}).
 The new configuration is accepted if the monomer does
not overlap with the rod.

We also considered a random walk on a square lattice (i.e., a bidimensional lattice).
 In this case the
starting configuration consists of a chain with a rescaled segment length
wound around the origin. In the initial configuration on the square
lattice one turn around the origin is performed by $8$ monomers. The
corner flip method is used here as well.

Finally, we used a freely jointed chain (FJC) to model three-dimensional
off-lattice polymers. 
Neighboring monomers have a fixed distance $a$ from each other and the rod has a
diameter $1.5\, a$ ensuring that monomers cannot accidentally pass from
one side to the other in the course of unwinding. The initial helical
configuration is such that one turn is performed with $10$ monomers.
Also in this case a time step of the dynamics
consists in $L$ attempts of moves for randomly chosen monomers. When a
monomer is selected, a new configuration is constructed via a rotation
of that monomer around the axis defined by the two neighboring monomers
by an angle randomly selected from $[0, 2\pi]$. The free end is updated
with a new random position preserving the bond length with the second
monomer. The new configuration is accepted if it does not overlap with
the rod.

For the case of a polymer with excluded volume we use only the model of a self-avoiding
walk (SAW) on the FCC lattice. The procedure is the same as for the ideal
chain on the FCC lattice, with the added constraint of excluded volume
between the monomers: one rejects moves violating it.

\begin{figure*}[!ht]
\hbox{
\includegraphics[angle=0,width=0.3575\textwidth]{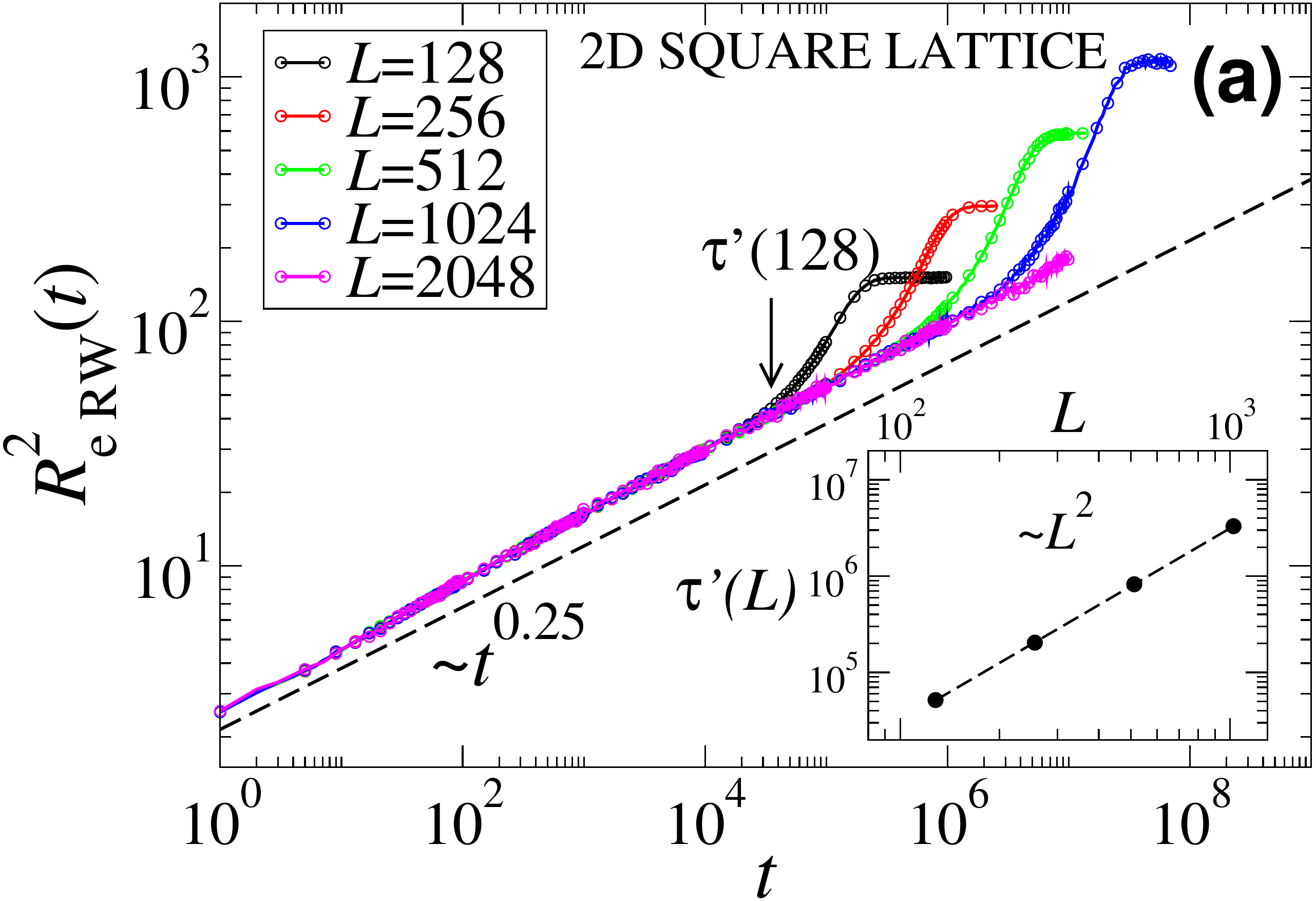}
\includegraphics[angle=0,width=0.31\textwidth]{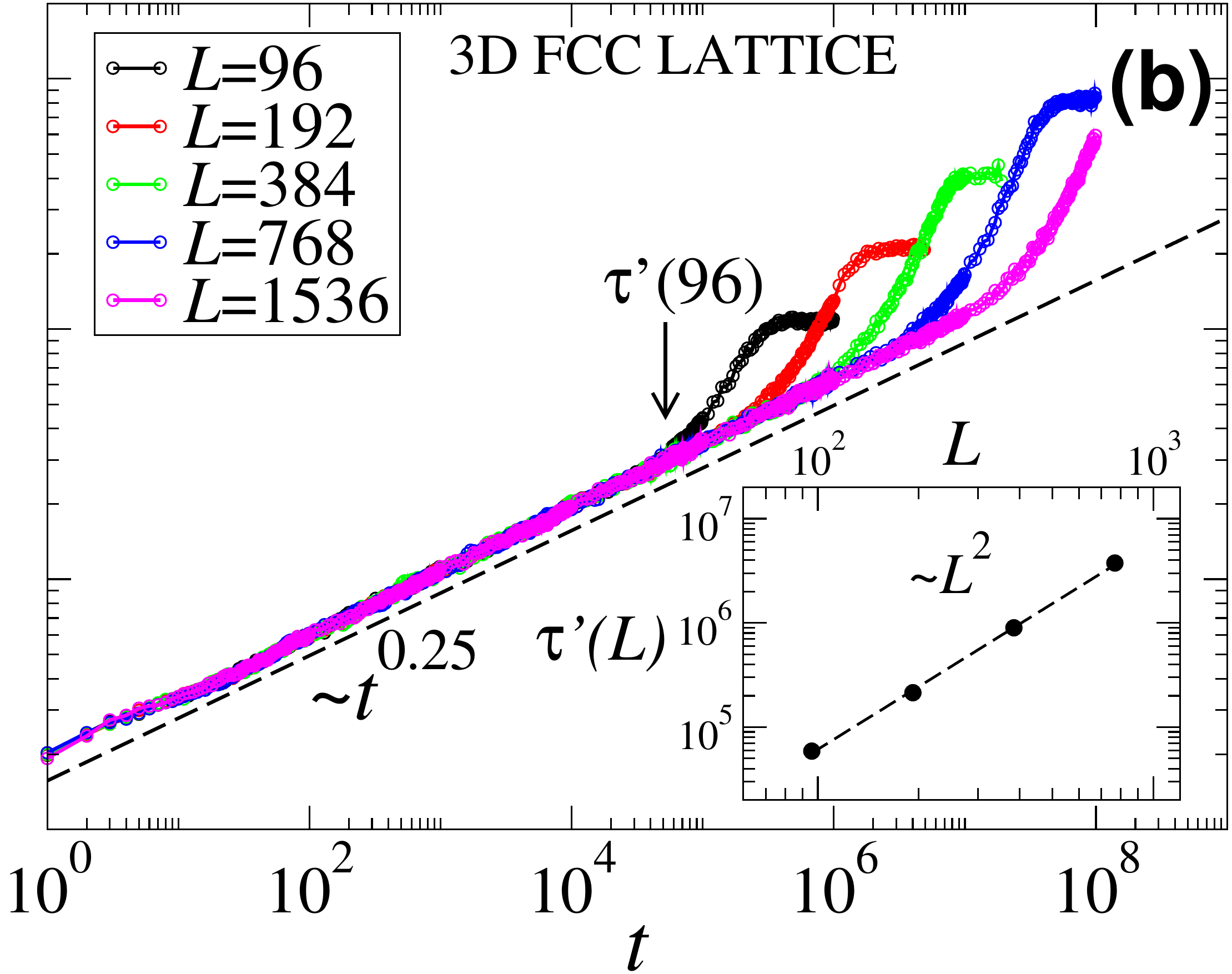}
\includegraphics[angle=0,width=0.31\textwidth]{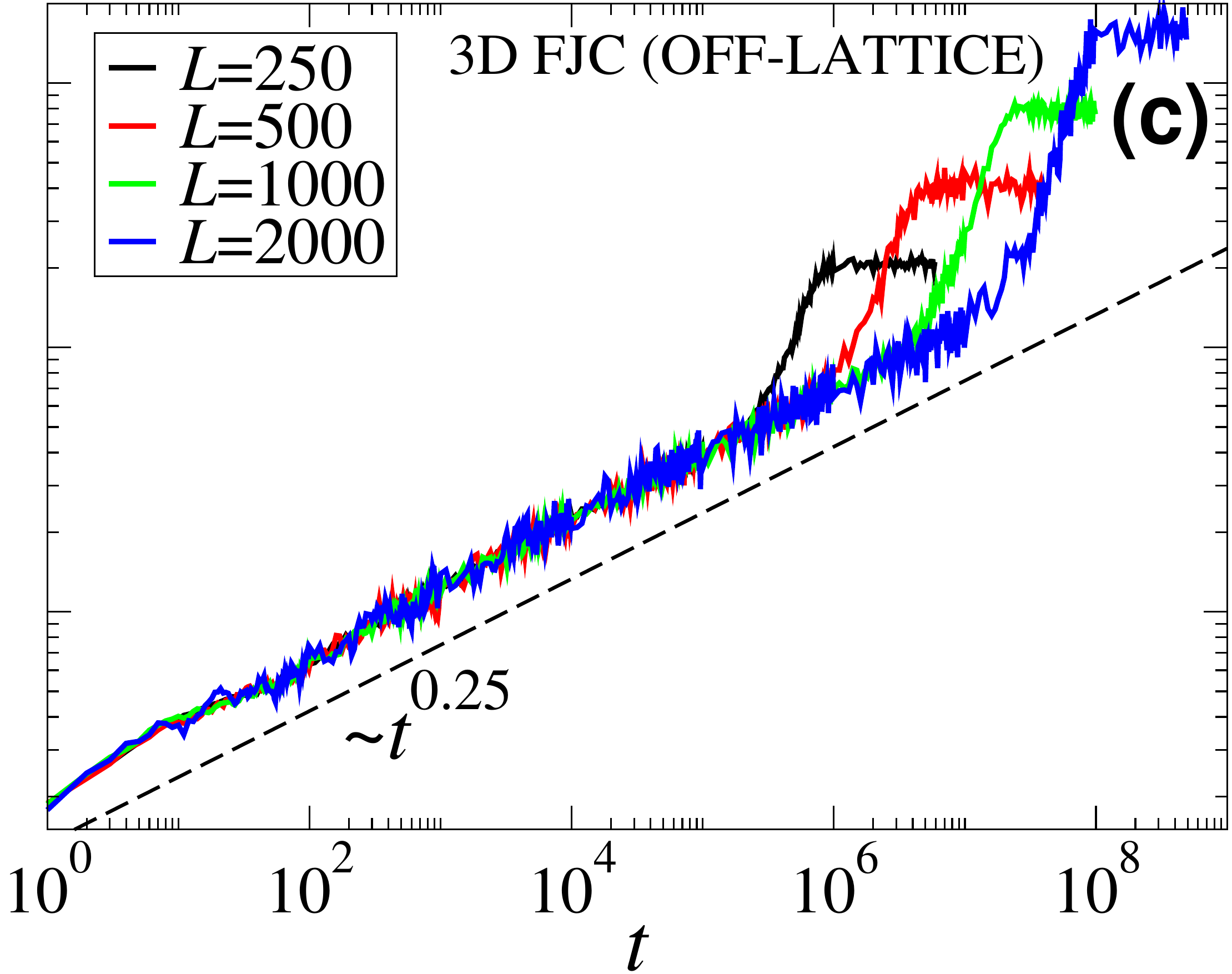}
}
\caption{Average squared distance $R_{\rm e\,RW}^2$ of the free
end from the rod versus time (ideal chain) for (a) a 2D polymer on a
square lattice, (b) a 3D polymer on a FCC lattice and (c) a FJC in 3D,
for various chain lengths. For the three models the short time regime
follows the power law $R_{\rm e\,RW}^2\sim t^{0.25}$. This regime ends
at times scaling as $\tau' \sim L^2$, see insets (the values of $\tau'$
for the shortest chains are indicated by vertical arrows).  The plateaus
indicate the completion of relaxation.  The averages
are made at least (for the largest sizes) over (a) $2000$ and (b)
$3000$ configurations. In (c), due to larger computational effort in
continuous space for a local move, the average is performed only over
$150$ configurations, which explains the larger noise.}
\label{fig:2}
\end{figure*}

\section{Results and Discussion}

\subsection{Short-Time Dynamics}

\label{sec:short}

\subsubsection{Radial Distance}

\begin{figure}[!ht]
\centering\includegraphics[angle=0,width=7cm]{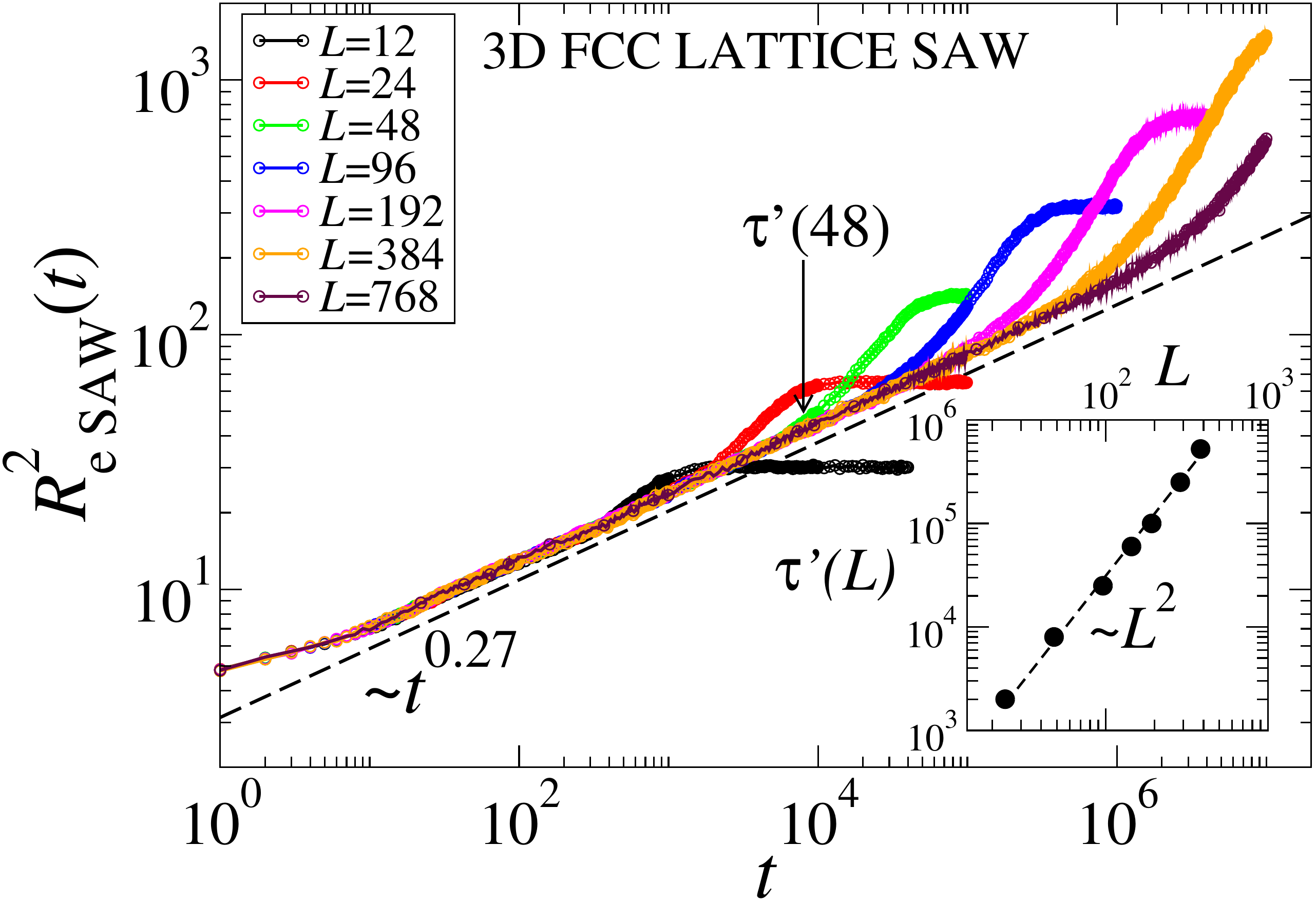}
\caption{Squared distance $R^2_{\rm e\,SAW}$ of the free end from the
rod versus time for a excluded volume chain on an FCC lattice. The short
time regime follows the power law $R^2_{\rm e\,SAW}\sim t^{0.27}$. The inset
shows the scaling of $\tau'$ with the polymer length. Averages
are made at least over $3000$ configurations (for the largest size).}
\label{fig:3}
\end{figure}

We consider first the ``radial" distance $R_{\rm e}$ which is defined as
the average distance of the free end monomer from the rod. This quantity
is shown in Figure~\ref{fig:2} (RW's) and~\ref{fig:3}
(SAW). Starting from a minimal value for the fully
wrapped configuration at $t=0$ the growth of $R_{\rm e}$ follows a power
law. In the ideal chain cases we find a first, short-time, regime:
\begin{equation}
R^2_{\rm e\,RW}(t)\sim t^{0.25}\,.
\label{R2RW}
\end{equation}
This scaling with time of a spatial length scale is remarkably slow
($|R_{\rm e}| \sim t^{1/8}$) compared to the usual relaxation time scales
encountered in polymer physics. The exponent is robust and is found
in the minimalistic 2D polymer on a square lattice (Figure~\ref{fig:2}(a)),
the RW on a FCC lattice (Figure~\ref{fig:2}(b)) and the freely jointed chain
off-lattice (Figure~\ref{fig:2}(c)).  We performed also simulations for a 3D
excluded volume chain on an FCC lattice, for which we find a power law
with a slightly larger exponent:
\begin{equation}
R^2_{\rm e\,SAW}(t)\sim t^{0.27}\,, 
\label{R2SAW}
\end{equation}
as it can be seen in Figure~\ref{fig:3}.  

\begin{figure}[!t]
\centering\includegraphics[angle=0,width=7cm]{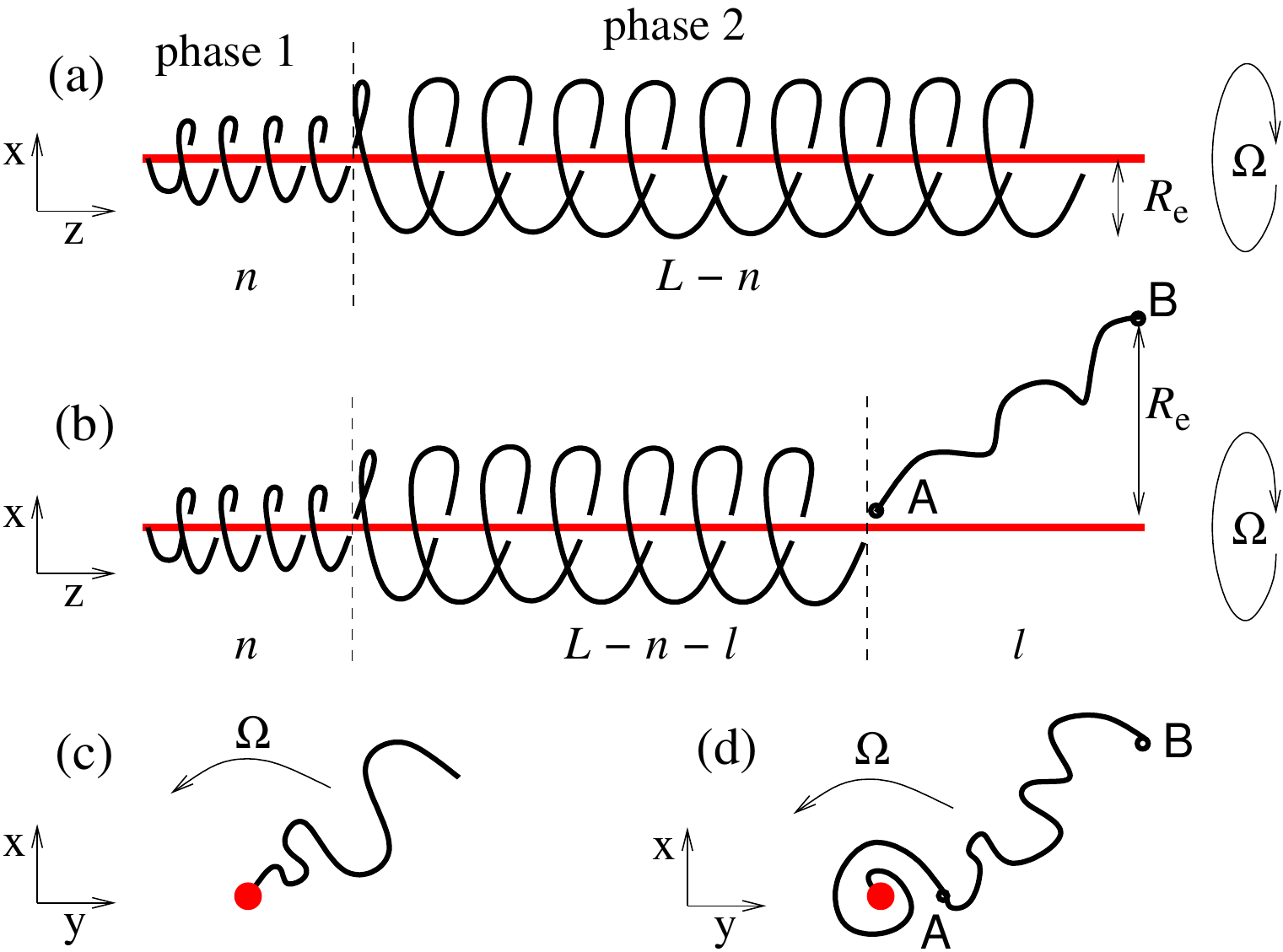}
\caption{(a-b) Configurations of the polymer during unwinding. (a) Two 
phase model consists of a tight helix (phase 1), which is the initial
conformation, and a loose helix with constant pace (phase 2). (b) Extension
of the two phase model accounting for the growth of an unwound coil
of length $l$ at the polymer end. (c-d) View of a polymer anchored 
to the rod (which is perpendicular to the plane) and rotating with 
angular velocity $\Omega$. If $\Omega$ is small the polymer rotates
while maintaining its equilibrium shape (case (c)). For high $\Omega$
the polymer gets partially wrapped around the rod, while a part of
length $l$ maintains its equilibrium shape. Equation~(\ref{length}) 
gives an estimate of the length $l$.}
\label{fig:config}
\end{figure}

To understand the origin of this exponent we model the unwinding starting
from a two phase picture.  We assume that during the early stage of the dynamics
the polymer starting from the fixed end has $n$ monomers tightly wrapped
around the rod, which are frozen as in their initial $t=0$ configuration,
while $L-n$ monomers are loose (we indicate these as phase 1 and phase 2,
respectively, see Figure~\ref{fig:config}(a)). We assume
that the loose monomers form a homogeneous helix with a constant pace,
but which is loosely wrapped around the rod. Obviously, if the loose helix
would extend until the free end of the polymer, the radial distance $R_{\rm e}$
would not grow in time. The two phase model is the starting point of our
analysis and we will focus on the configuration of the polymer close to
the free end later. We assume that the dynamics is governed by the
following equation:
\begin{equation}
\gamma_2 \frac{dn}{dt} = - \frac{\partial {\cal F}}{\partial n}\,,
\label{eq:domain1}
\end{equation}
which is a balance between frictional and ``entropic'' forces during the
growth of the helical domain.  Here ${\cal F}(n) = f_1 n + f_2
(L- n)$ is the total free energy of the configuration, with $f_1$ and
$f_2$ the free energies per monomer of the two phases (with $f_2 < f_1$
as the loose helix has a higher entropy than the tight helix).
$\gamma_2$ is the friction coefficient. The total
winding angle of the last monomer is equal to $2\pi$ times the number
of times the polymer is wrapped around the rod.  For the two helices
model such a quantity is then given by:
\begin{equation}
\Theta = n \Delta \theta_1 + (L-n) \Delta \theta_2\,,
\label{thetan}
\end{equation}
where $\Delta \theta_1$ and $\Delta \theta_2$ are the densities of winding
for the two phases (with $\Delta \theta_1 > \Delta \theta_2$,
as phase 2 is more loosely wrapped compared to phase 1).  A decrease
in $n$ leads to a decrease in the total winding angle and the whole
loose helix rotates in a corkscrew motion. The friction coefficient is
then proportional to the length of the rotating domain $\gamma_2 \sim
(L- n)$. Using this input and the form of ${\cal F}(n)$ we get from
the integration of eq.~(\ref{eq:domain1}):
\begin{equation}
L-n \sim \sqrt{t}\,.
\label{growth1}
\end{equation}
The angular velocity is given by eq.~(\ref{thetan}):
\begin{equation}
\Omega = \frac{d\Theta}{dt} \sim \frac{dn}{dt} \sim \frac{1}{\sqrt{t}}\,,
\label{omega}
\end{equation}
which decreases in time as there is an increasing friction when
the loose helix grows. 

The assumption that phase 2 is a helix of constant
pace and radius is an approximation. However, as we will show,
the numerical data are in good agreement with a square root growth
(eq.~(\ref{growth1})), which is a consequence of that assumption.  Note that
the assumption can be relaxed, allowing phase 2 to have fluctuations;
the only essential requirement is that the friction $\gamma_2$ should
scale linearly with the length of the domain. Our approximation is similar in spirit to the monoblock
approximation where an inhomogeneously stretched polymer is modeled
by a homogeneously stretched one, an approximation that does not change the
scaling of the large scale geometry of the deformed chain~\cite{brochard94,schiessel97}.

Let us consider now the growth of a coiled domain at the end of the
chain (the stretch AB of length $l$ shown in Figure~\ref{fig:config}(b)).
This domain grows from a polymer rotating with angular velocity
$\Omega$. Let us assume that the friction originating from  the
coiled part is negligible compared to that of the loose helix, so the
calculations leading to eqs.~(\ref{growth1}) and (\ref{omega}) remain
valid. To understand the coil growth we consider a polymer attached to
a rod and rotating with angular velocity $\Omega$. If the polymer is
sufficiently short and $\Omega$ small, its equilibrium conformation is
not perturbed by the rotation and in particular it will have no winding
(Figure~\ref{fig:config}(c)). If the polymer length exceeds a given threshold
value, then part of the polymer close to the attachment point gets wound
while the final part rotates maintaining its equilibrium shape (length
of the part AB in Figure~\ref{fig:config}(d)).  We estimate now the length of
the end coil for a rotating polymer. To understand the calculation it is
useful to consider the analogous problem of a polymer pulled by one end
by a constant force~\cite{brochard93,brochard94,schiessel95,rowg12,sakaue12}.
The polymer maintains its equilibrium conformation if the applied force,
$f$, or the polymer length, $l$, do not exceed the values fixed by
the equation:
\begin{equation}
f R_F \sim k_B T\,,
\label{fRF}
\end{equation}
where $R_F \sim l^\nu$ is the Flory radius, 
$k_B$ the Boltzmann's constant and $T$ the temperature. For a polymer
rotating with an angular velocity $\Omega$ the force which distorts its
shape is due to friction. The expression analogous to eq.~(\ref{fRF})
is then given by
\begin{equation}
\gamma v R_F \sim  k_B T\,.
\label{angular_fRF}
\end{equation}
Using $v = \Omega R_F$ and $\gamma \sim l$ for a Rouse polymer we obtain
the following relation for the length of the coiled unwound end of the
rotating polymer:
\begin{equation}
l^{1+2\nu} \sim  \frac{k_B T}{\Omega}.
\label{length}
\end{equation}
Using the law (\ref{omega}) for the angular velocity we finally obtain
for the growth of $l$:
\begin{equation}
l \sim t^{1/(4\nu+2)}\,,
\label{evol_length}
\end{equation}
and, from the equilibrium relation $R_{\rm e}^2 \sim l^{2\nu}$ we find that
the squared distance from the rod grows as:
\begin{equation}
R_{\rm e}^2 \sim t^z\,, 
\label{Revst}
\end{equation}
with $z=\nu/(2\nu+1)$. Note that $z=1/4$ for a Gaussian polymer
($\nu=1/2$) and $z\simeq 0.27$ for a self-avoiding polymer ($\nu \simeq
0.59$) which is in excellent agreement with the numerical results
(Figures~\ref{fig:2} and~\ref{fig:3}).

\begin{figure*}[!ht]
\hbox{
\includegraphics[angle=0,width=0.34\textwidth]{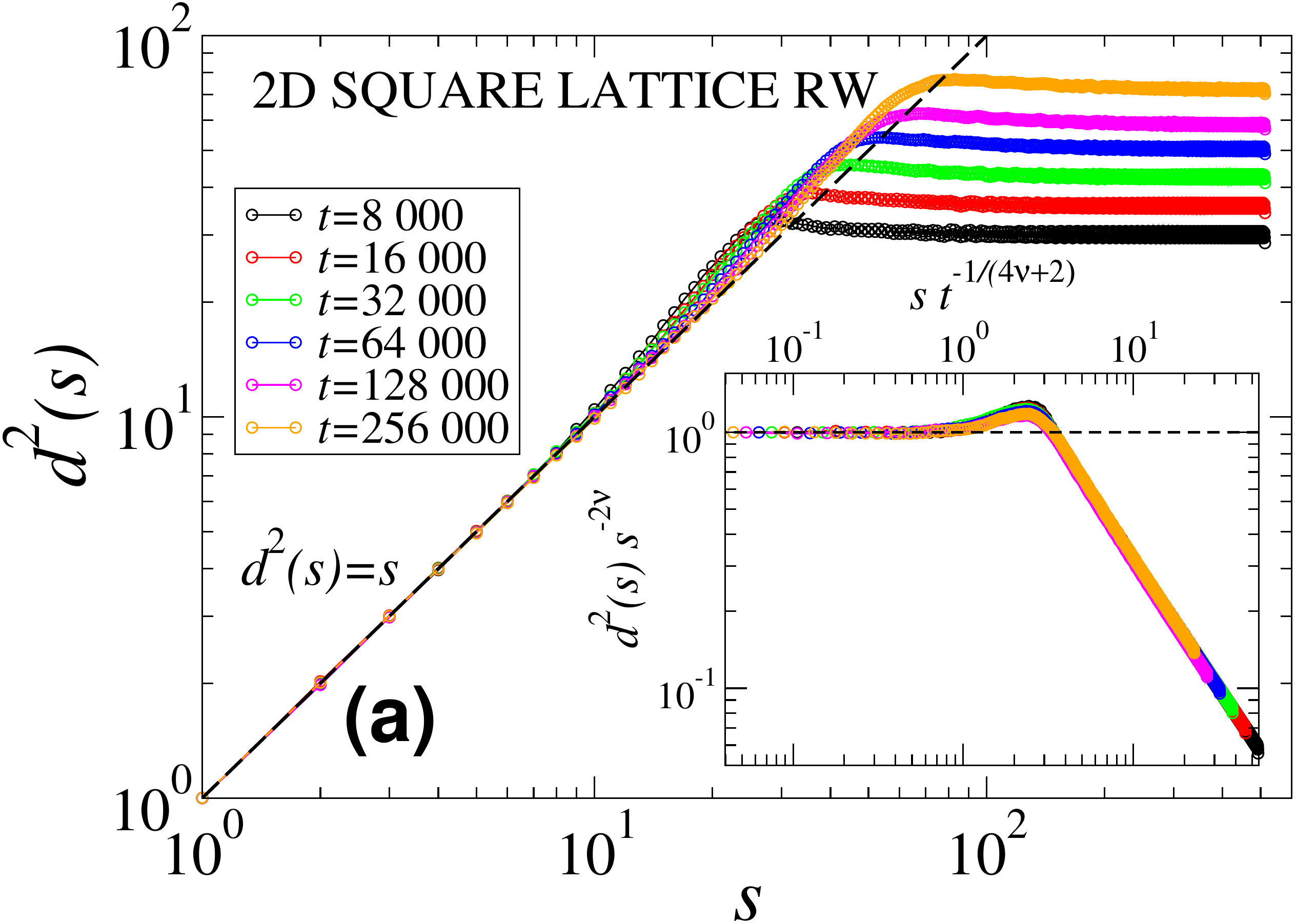}
\includegraphics[angle=0,width=0.325\textwidth]{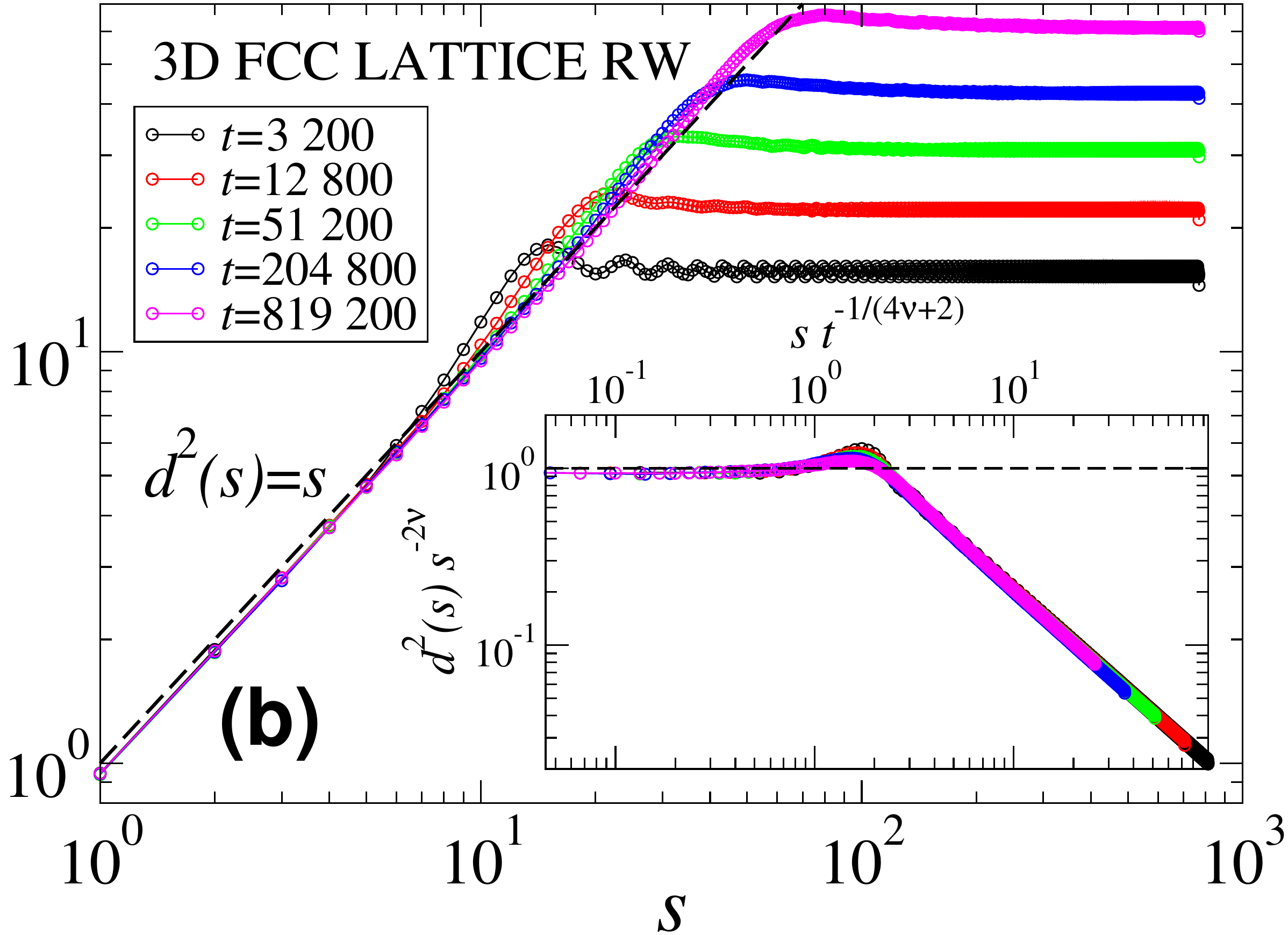}
\includegraphics[angle=0,width=0.315\textwidth]{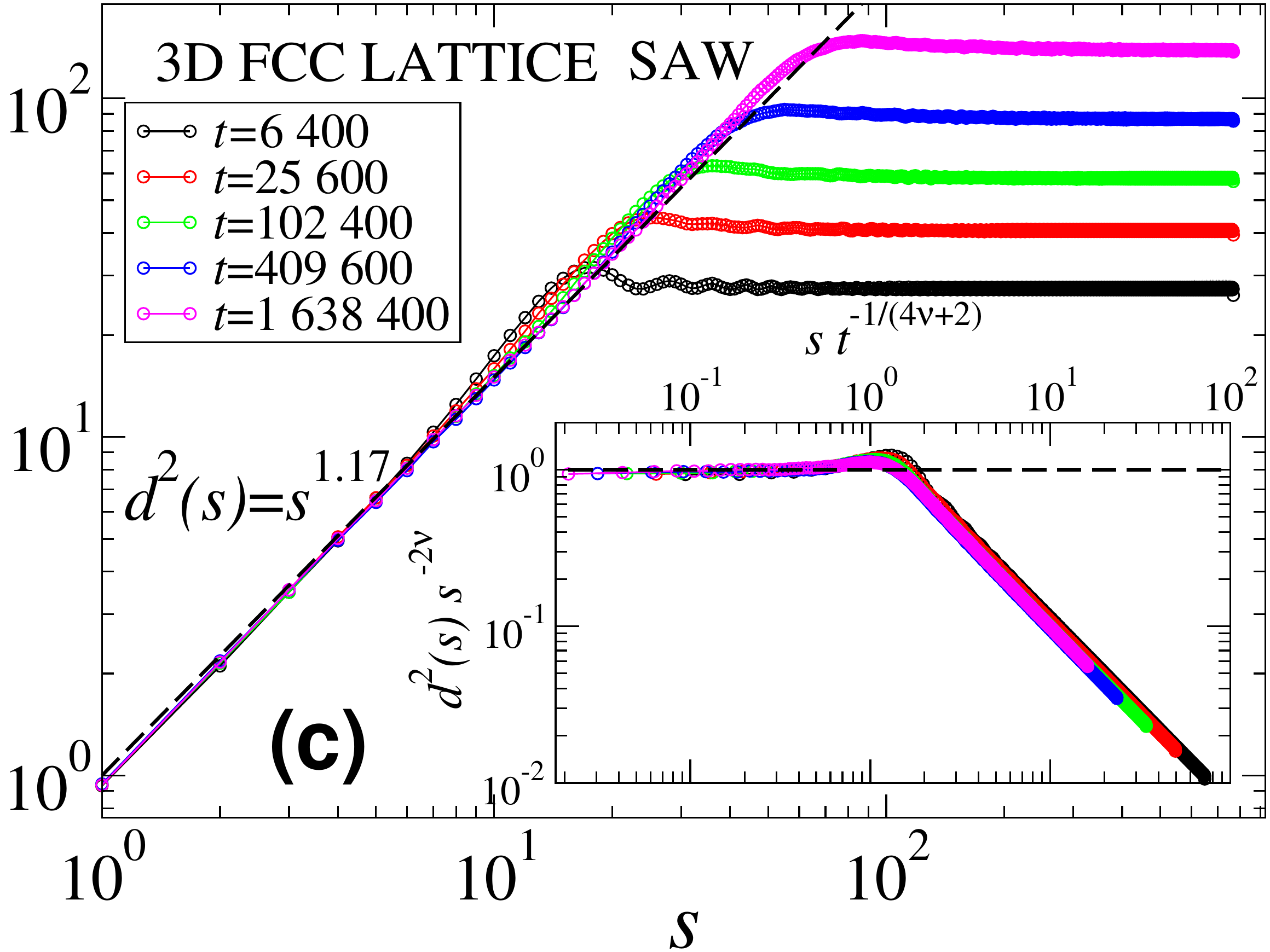}
}
\caption{Plot of $d^2(s)$ vs. $s$, for different times in the short time regime
and $L=768$. The quantity $d(s)$, defined in eq.~(\ref{def_ds}), is
the planar distance of the last monomer (index $L$) at the free end from
the $(L-s)$-th monomer. For small $s$, one has $d^2(s) \sim s^{2\nu}$,
indicating that the polymer forms a random coiled phase from its free end.
The region where $d^2(s) \sim s^{2\nu}$ holds grows in time.
The cases shown
are (a) 2D RW, (b) RW on the FCC lattice, (c) SAW on the FCC lattice.
The insets show rescalings of the data, see eq.~(\ref{scaling_ansatz}).
}
\label{fig:4}
\end{figure*}

We expect that the above description remains valid until the loose helix
has grown to reach the first monomer. This corresponds to $n=0$,
i.e. when the tight helix has disappeared. According
to eq.~(\ref{growth1}) this happens at a characteristic time $\tau'$
scaling as $\tau' \sim L^2$. We estimated $\tau'$ for polymers of
different lengths from the simulation data of the radial distance
of Figures~\ref{fig:2} and \ref{fig:3}. This is the time at which the growth
law starts deviating from eq.~(\ref{Revst}).  The vertical arrows in Figure~\ref{fig:2}
mark the estimated $\tau'$ for the polymer of shortest length.
The insets of Figures~\ref{fig:2}(a) and (b) show plots of
$\tau'$ vs. $L$ as obtained from the data of the main plots. There is
an excellent agreement with the predicted scaling $\tau' \sim L^2$. An
equally good agreement was found for all the other cases studied. Note
that the scaling $\tau' \sim L^2$ does not depend on the presence of
self-avoidance, as demonstrated by the data in the inset of Figure~\ref{fig:3}.

\begin{figure*}[!t]
\hbox{
\includegraphics[angle=0,height=4.1cm]{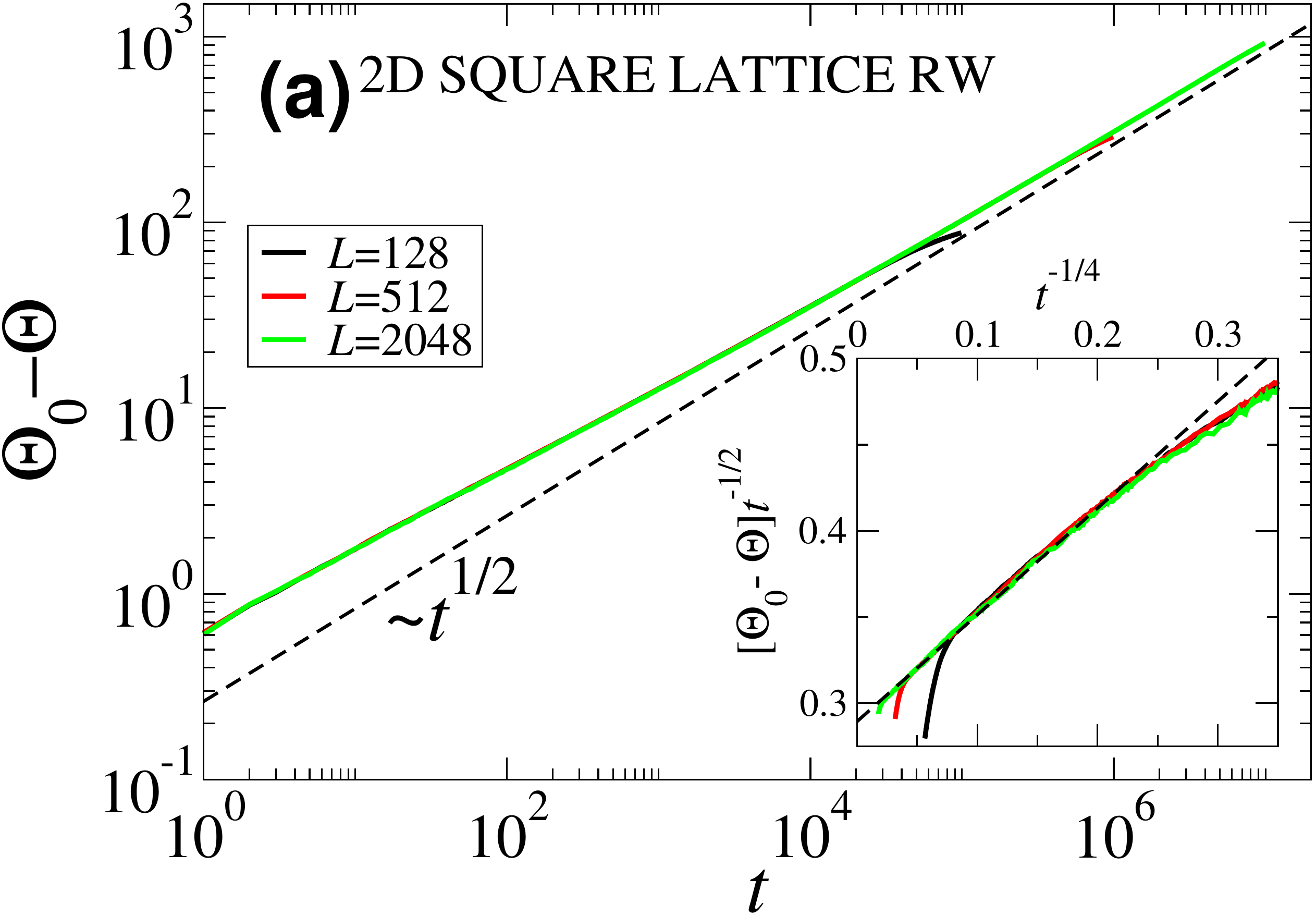}
\includegraphics[angle=0,height=4.1cm]{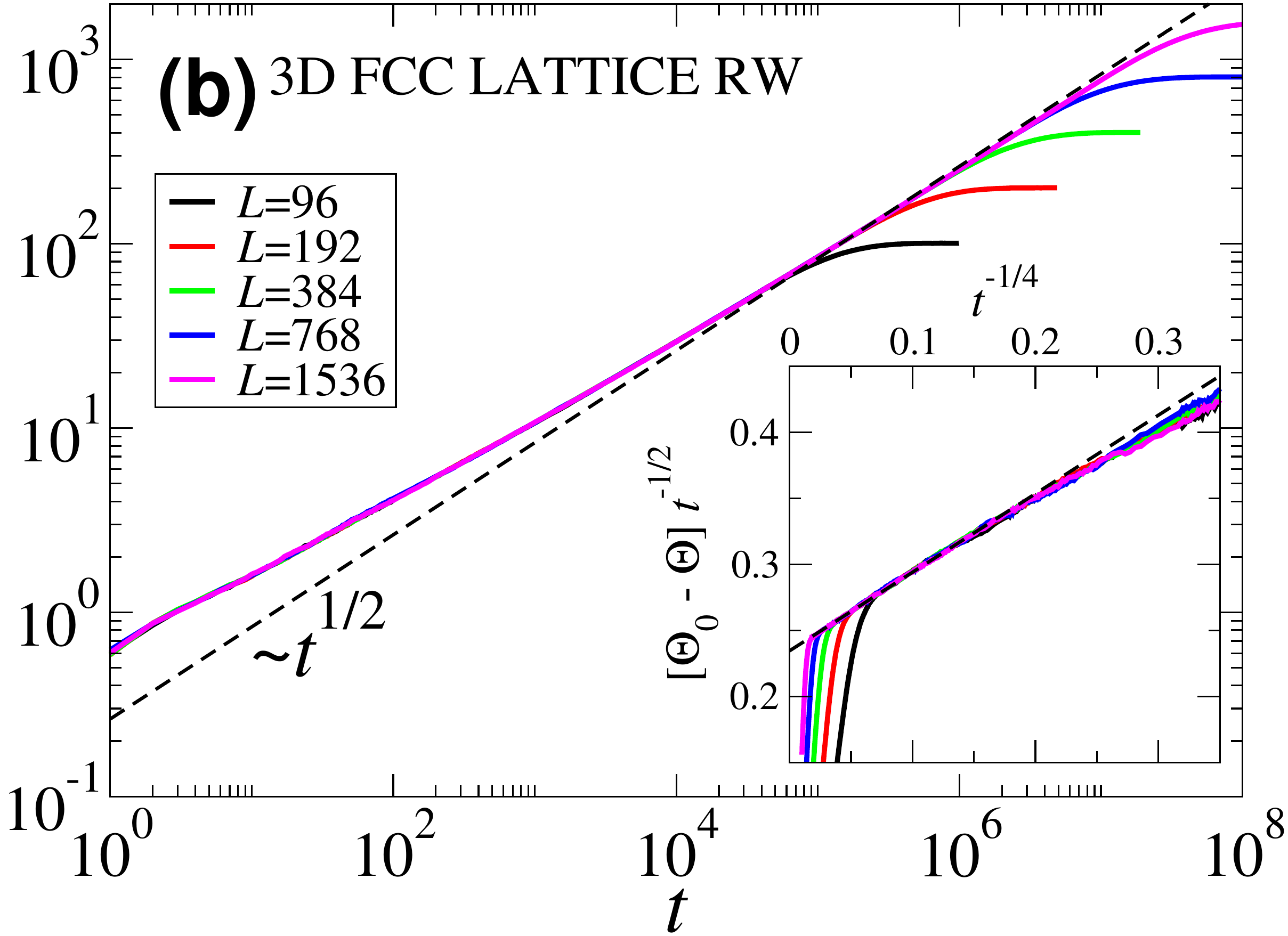}
\includegraphics[angle=0,height=4.2cm]{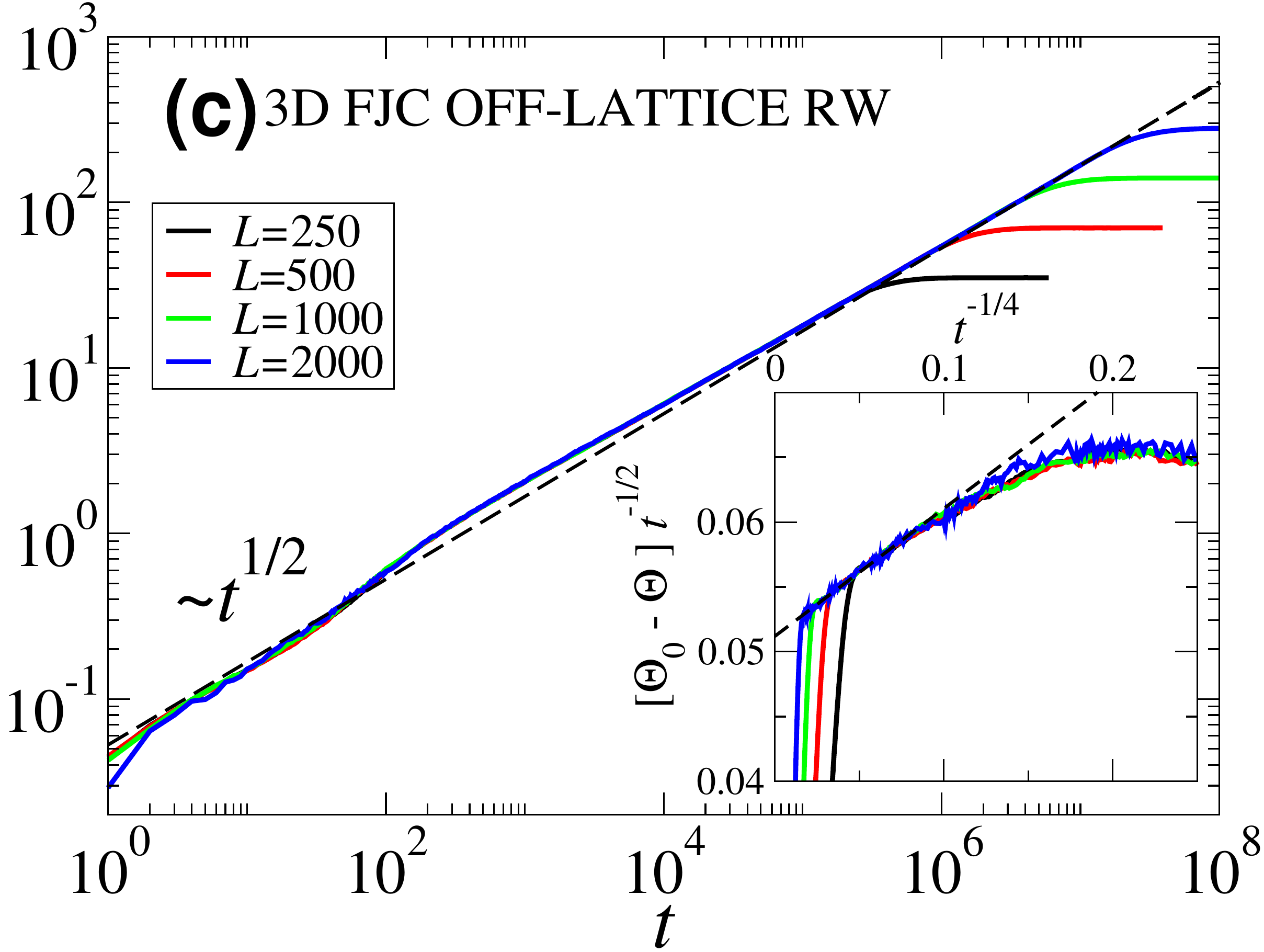}
}
\caption{Scaling of $\Theta_0-\Theta$ vs. time, for different polymer lengths, for the ideal chain models. 
The numerics are well fitted by a power law $\sim t^{1/2}$. (Insets) Plots of $(\Theta_0 - \Theta)t^{-1/2}$ vs. $t^{-1/4}$.
The dashed line is the slope predicted from eq.~(\ref{corr_scaling}).
}
\label{short_theta}
\end{figure*}

We consider next the distance $d(s)$ between the free end monomer at
position $L$ and a monomer $L-s$ projected onto a plane perpendicular
to the rod.  In two dimensional models this is equal to the total
distance $d(s) \equiv |\vec{R}_L-\vec{R}_{L-s}|$ where $\vec{R}_i$
is the position of monomer $i$. In three dimensions, for a rod parallel
to the $z$-axis we have:
\begin{equation}
d(s) = \sqrt{\left(x_L-x_{L-s}\right)^2 + \left(y_L-y_{L-s}\right)^2}\,,
\label{def_ds}
\end{equation}
where $x_i$ and $y_i$ are the coordinates of the monomer $i$. In all
cases analyzed (see Figure~\ref{fig:4}) we find for small $s$ a scaling $d^2(s)
\sim s^{2\nu}$ which demonstrates that the end part of the polymer is an
equilibrated coil. The plots show that $d^2(s)$ saturates at a constant
value for sufficiently large value of $s$. This point identifies the end
of the coil and the beginning of the loose helical region. From scaling
arguments we expect:
\begin{equation}
d^2(s) = s^{2\nu} g\left(\frac {s}{t^{1/(4\nu+2)}}\right)\,,
\label{scaling_ansatz}
\end{equation}
where for small values of $x=s/t^{1/(4\nu+2)}$ the function $g(x)$
converges to a constant. For large $x$ we expect $g(x)\sim 1/x^{2\nu}$
as $d^2(s)$ is independent of $s$, as the projected distance from the
end monomer to the monomers in the helical domain cannot increase.
The inset of Figure~\ref{fig:4} shows a scaling plot of $d^2(s) s^{-2\nu}$
vs. $s/t^{1/(4\nu+2)}$ in full agreement with the scaling ansatz
(\ref{scaling_ansatz}). This result supports the prediction that
the coil growths according to eq.~(\ref{evol_length}). Note that the
scaling function has a maximum at the crossover between the two scaling
regimes. The maximum is not very pronounced but indicates that the
polymer, compared to an equilibrated coil, is slightly more stretched
in the vicinity of the rod.

\subsubsection{Dynamics of the Total Winding Angle} 

\begin{figure}[!bt]
\hbox{
\includegraphics[angle=0,width=7.8cm]{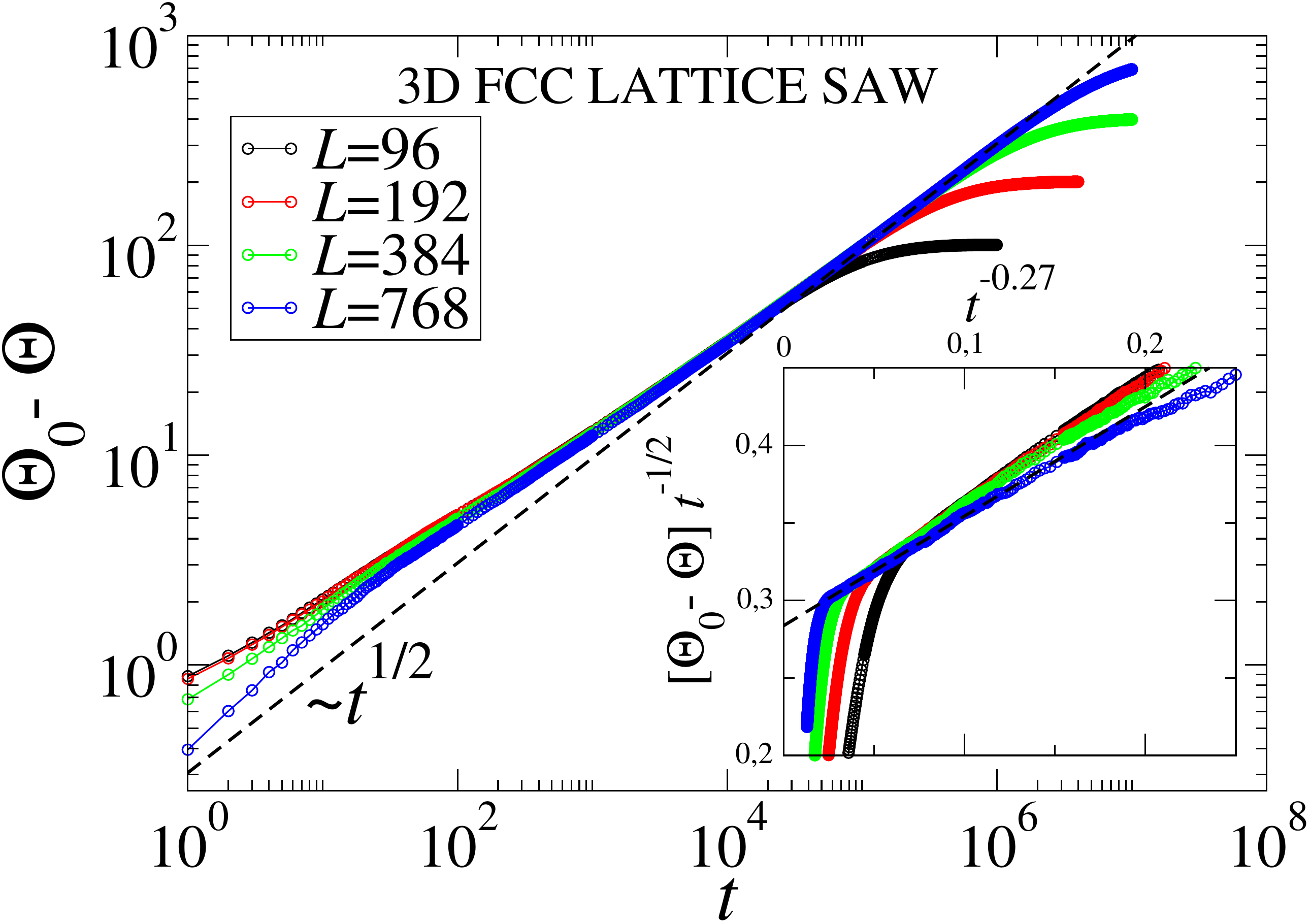}
}
\caption{Scaling of $\Theta_0-\Theta$ vs. time, for different polymer lengths, for the SAW model. 
(Inset) Plot of $(\Theta_0 - \Theta)t^{-1/2}$ vs. $t^{-0.27}$. The dashed line is the slope predicted
from eq.~(\ref{corr_scaling}).
}
\label{short_theta_SAW}
\end{figure}

The theory developed in the previous section can be tested also
on the dynamics of the total winding angle. Taking into account the
presence of the coil of length $l$ at the polymer end we need to modify
eq.~(\ref{thetan}) with:
\begin{equation}
\Theta = n \Delta \theta_1 + (L-l-n) \Delta \theta_2\,,
\label{thetan2}
\end{equation}
as the coil, which is of length $l$, does not contribute to the
winding. We define with $\Theta_0 = L \Delta \theta_1$ the initial
total winding angle. Combining eqs.~(\ref{growth1}), 
(\ref{evol_length}) and (\ref{thetan2}), we then get:
\begin{eqnarray}
\Theta_0-\Theta &=& (L-n) (\Delta \theta_1 - \Delta \theta_2 ) 
+ l \Delta \theta_2\,, \nonumber\\
 &=& A t^{1/2} + B t^{1/(4\nu+2)}\,, \nonumber\\
 &=& A t^{1/2} \left( 1 + \frac{B}{A} \frac{1}{t^{\nu/(2\nu+1)}} \right)\,,
\label{corr_scaling}
\end{eqnarray}
with $A$ and $B$ being some positive constants. The prediction is that
$\Theta_0 - \Theta$ scales as $\sim \sqrt{t}$, with a slowly
decaying correction term originating from the
equilibrated end coil. The correction is predicted to scale as $\sim
t^{-1/4}$ for a RW and $\sim t^{-0.27}$ for a SAW.  In order to test
the validity of eq.~(\ref{corr_scaling}) we plot in Figure~\ref{short_theta}
the quantity $\Theta_0-\Theta$ as a function of $t$ in a log-log scale
for three ideal-chain  models in two and three dimensions, and the same
for SAW's in Figure~\ref{short_theta_SAW}.  In all cases the data approach for
sufficiently long times the expected $\sim \sqrt{t}$ law.

To investigate the nature of the corrections to the leading scaling
behavior we plot in the insets of Figure~\ref{short_theta} the quantity
$(\Theta_0-\Theta)/\sqrt{t}$ vs. $t^{-\nu/(2\nu+1)}$. The data for short
times and for different polymer lengths follow a straight line in good
agreement with the prediction of eq.~(\ref{corr_scaling}). The slope
of the lines are positive and imply $B > 0$, as expected. Some stronger
nonmonotonic behavior is observed in the 3D off-lattice model which does
not have a counterpart in the other cases studied. The behavior of the
winding angle was also investigated in a previous publication~\cite{wal13}
and estimated to scale in the early time dynamics as $\Theta_0-\Theta\sim
t^{\rho}$ where $\rho\approx 0.43-0.44$. This seemed to match the short
time dynamics rather well, although a closer inspection of the data shows
that eq.~(\ref{corr_scaling}) fits the data better. The analysis of
the local winding, which follows, gives further support of a $\sqrt{t}$
growth of the unwound domain.

\subsubsection{Local winding angle} 

\begin{figure}[!bt]
\includegraphics[angle=0,width=0.45\textwidth]{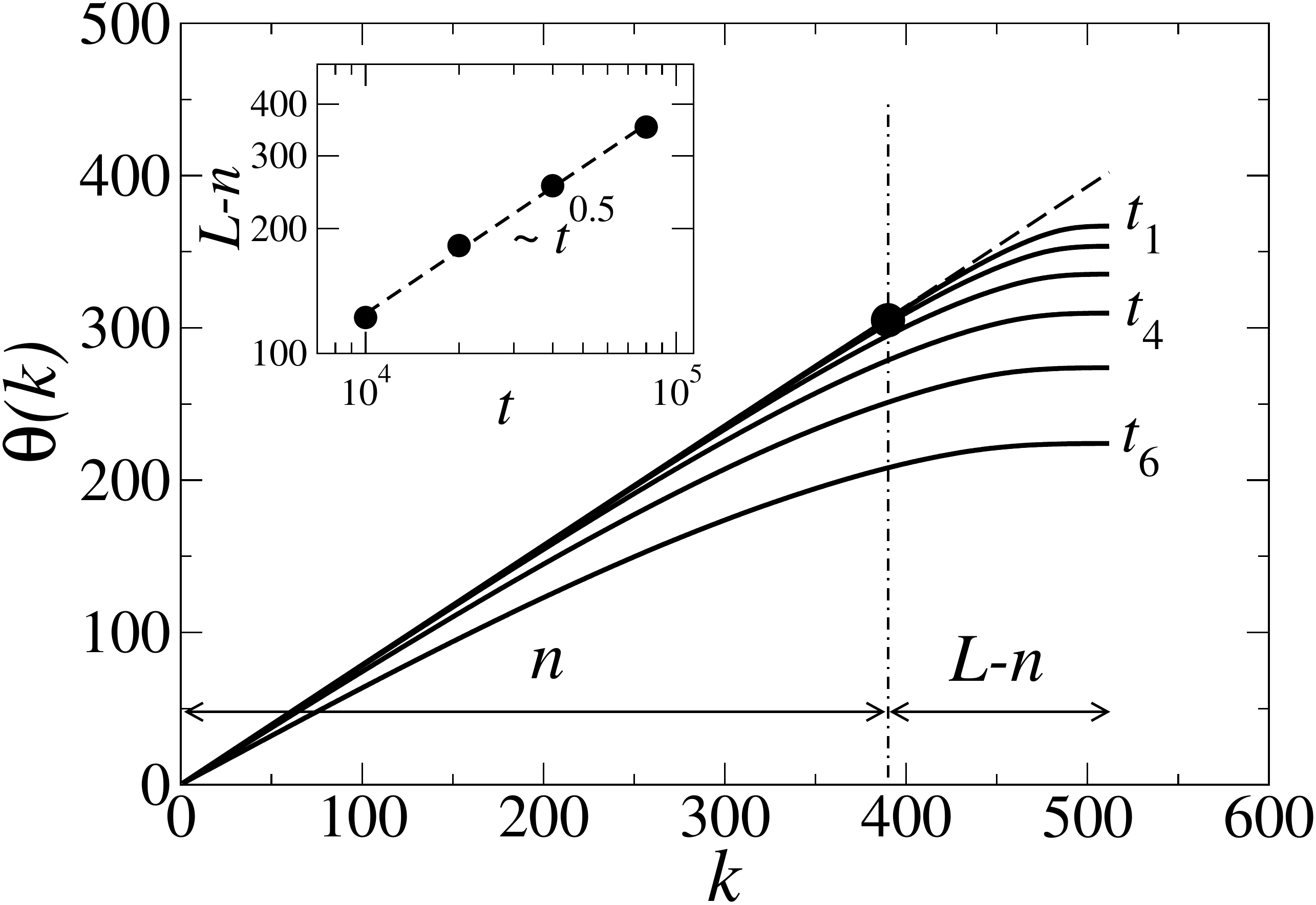}
\caption{Time evolution of the local winding angle $\theta(k)$ vs. 
monomer index $k$ for the 2D RW of size $L=512$. The dashed tilted line corresponds to the 
fully wrapped conformation $\theta(k) =\Delta \theta_1 k$. The 
data are obtained for increasing times $t_1=10\,000$, $t_2=20\,000$,
$t_3=40\,000$, $t_4=80\,000$, $t_5=160\,000$ and $t_6=320\,000$.
The vertical dashed-dotted line denotes the boundary between the tight
helix of length $n$ and the loose part of length $L-n$ at the time
$t_1$. Inset: The growth of the loose domain follows the square root
behavior predicted by eq.~(\ref{growth1}).}
\label{fig:local_winding}
\end{figure}

Further insight of the polymer dynamics can be obtained from the analysis
of the local winding angle $\theta(k)$, which is the winding angle of the
$k$-th momomer. As the winding angle is counted from the monomer attached
to the rod ($k=1$) one has $\theta(1)=0$, whereas the total winding angle
defined above is $\Theta = \theta(L)$. Figure~\ref{fig:local_winding}
shows the time evolution of $\theta(k)$ vs. $k$ for different times
and for a polymer of length $L=512$ for a planar RW.  At $t=0$ the
configuration is fully wrapped around the rod which corresponds to a
linear increase $\theta(k) = \Delta \theta_1 k$.  As the time evolves
$\theta(k)$ decreases in a more pronounced manner starting from the
free end of the polymer; at short times there is a domain of length $n$
which is still fully wrapped as at time $t=0$, followed by a loose part
of length $L-n$.  The inset of Figure~\ref{fig:local_winding} shows a plot
of $L-n$ vs. $t$. The data are in very good agreement with the square
root growth predicted by eq.~(\ref{growth1}). Differently from the data
for the total winding angle of eq.~(\ref{corr_scaling}), in this case
there are no corrections to scaling expected.

\subsection{Late Stage Relaxation} 
\label{sec:late}

\begin{figure*}[!ht]
\hbox{
\includegraphics[angle=0,width=0.3525\textwidth]{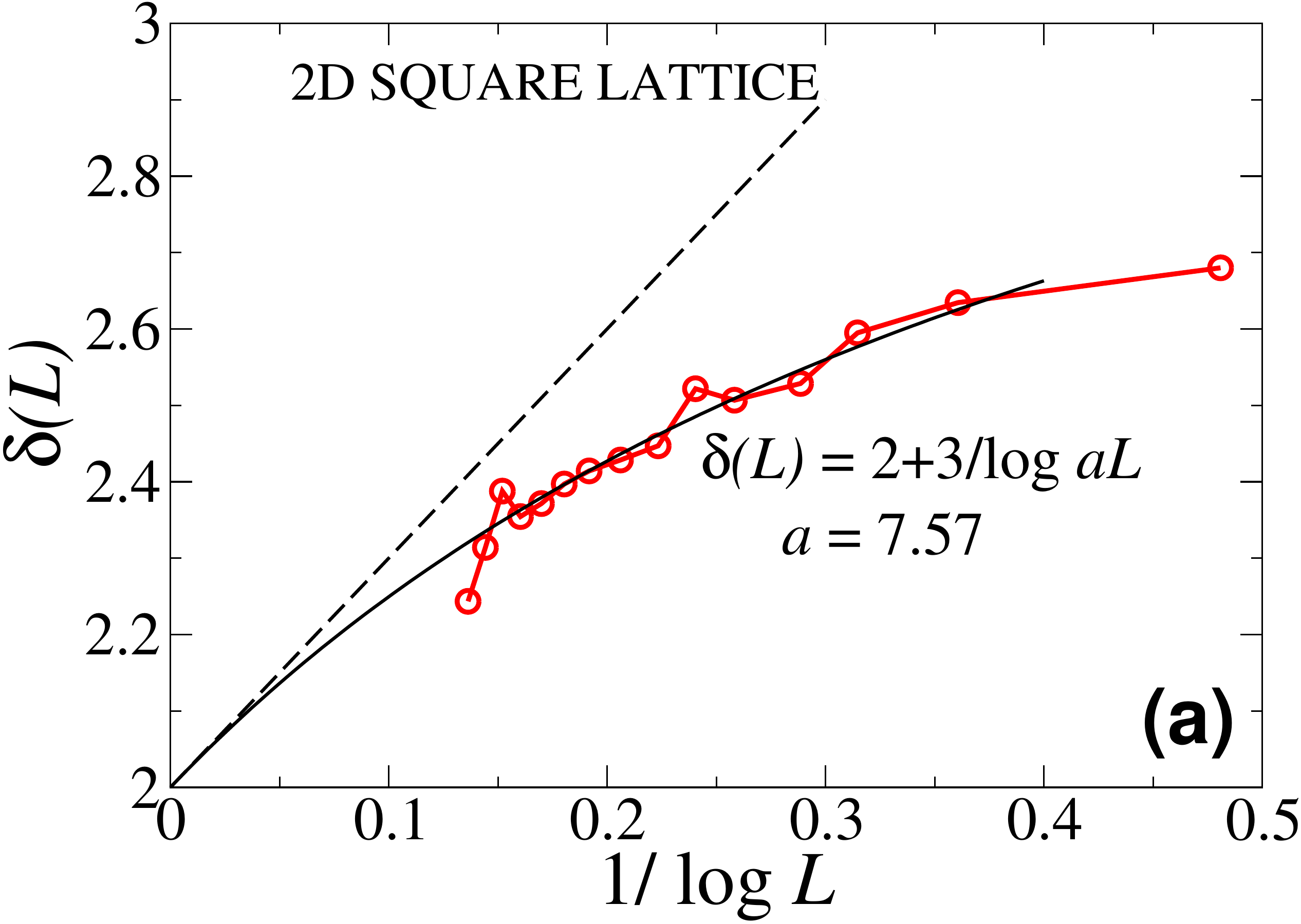}
\includegraphics[angle=0,width=0.31\textwidth]{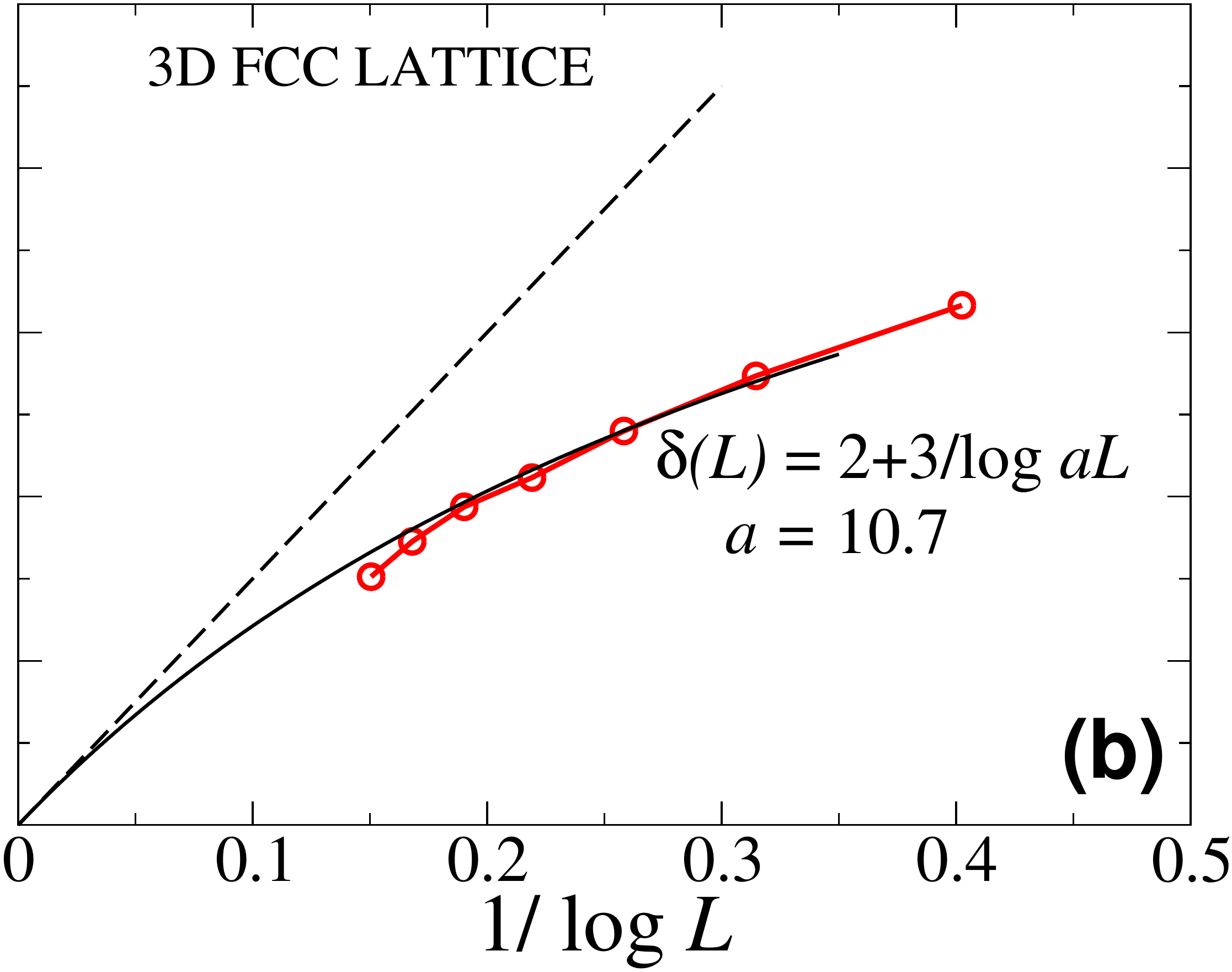}
\includegraphics[angle=0,width=0.31\textwidth]{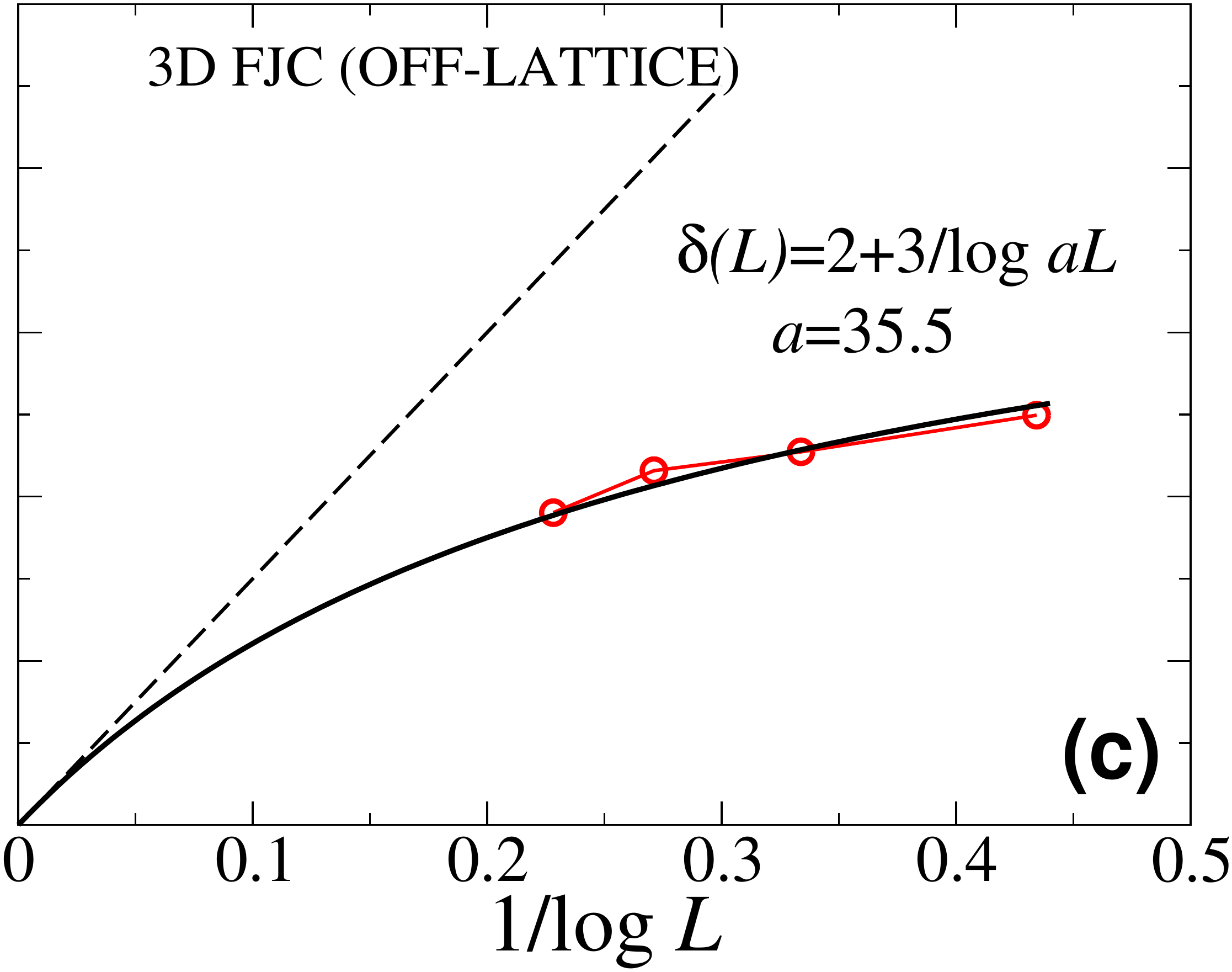}
}
\caption{Effective exponent defined by eq.~(\ref{eff_exp}) for models without
self-avoidance in two and three dimensions. Asymptotically in $L$
this quantity is expected to converge to $2\nu+1=2$. The solid line is given by
eq.~(\ref{eff_exp}) with a single adjustable parameter, the
scale factor $a$.}
\label{fig:eff_exp}
\end{figure*}

In the previous section we have fully characterized the early time
relaxation dynamics. We found that the first regime in which the typical
polymer configuration looks like in Figure~\ref{fig:config}(b) ends at a time
scaling as $\tau' \sim L^2$. At very long times the dynamics was studied
in ref.~\cite{wal13} using a force balance equation for the total winding
angle $\Theta$. This equation reads:
\begin{equation}
\gamma_\tau \frac{d\Theta}{dt} = - \frac{\partial {\cal F}}{\partial\Theta}\,,
\label{late_stage_eq}
\end{equation}
where ${\cal F}$ is the free energy of a polymer in equilibrium with
a total winding angle $\Theta$ and where $\gamma_\tau\sim L^{1+2\nu}$
is the friction coefficient. The free energy is a function of a scaling
variable $\Theta/(\log L)^\alpha$, where $\alpha$ is an exponent governing
the fluctuations of the winding angle at equilibrium. For RW's it is known
rigorously~\cite{rudnick88} that $\alpha=1$ while numerical simulations
of 3D SAW's yield $\alpha\simeq 0.75$~\cite{walter11}.  For small winding
angles the free energy is quadratic in the scaling variable, so that the
relaxation to equilibrium becomes exponential~\cite{wal13}: $\Theta(t)
\sim e^{-t/\tau}$ where $\tau$ is the longest relaxation time:
\begin{equation}
\tau  \sim L^{2\nu+1}(\log L)^{2\alpha}\,.
\label{taup}
\end{equation} 
Since the initial configuration is fully wound with $\Theta_0 \sim
L$, the total unwinding time is given by $\tau^* \sim \tau \log L$
\cite{wal13}.  The data are best analyzed using the definition of
a running exponent, which probes the local slope of the data in a
log-log plot.  From eq.~(\ref{taup}) we get:
\begin{equation}
\delta(L)\equiv\frac{d\log\tau^*}{d\log L} =
1+2\nu+\frac{2\alpha+1}{\log (aL)}\,,
\label{eff_exp}
\end{equation}
where we have included a scale term $a$, in order to account for further
corrections to scaling.  Figure~\ref{fig:eff_exp} shows a plot of the
numerical value of $\delta (L)$ in two and three dimensions obtained
from simulations of RW's. These data extend those of ref.~\cite{wal13}
by including the three dimensional case for both FCC lattice (b) and
off-lattice (c) models. The data are compared with eq.~(\ref{eff_exp})
where there is only a single adjustable parameter $a$ used in the fit. The
agreement is very good confirming the validity of the analytical approach
of force balancing in eq.~(\ref{late_stage_eq}), which describes the
process using the total winding angle as a single reaction coordinate.

\section{Conclusion}

In this paper we have investigated the problem of the unwinding dynamics
of a flexible polymer from a rigid rod. Scaling arguments and force
balance equations allowed us to fully characterize the early stages
of the unwinding and the late stages of the relaxation dynamics. These
arguments are supported by extensive numerical simulations in two and
three dimensions, with and without self-avoidance.  The early dynamics can
be understood by a three phase picture, where the polymer configuration
starting from the fixed end can be described by a tight helix, a looser
helix and a free random coil. The latter two phases grow in time following
two different dynamical laws as predicted by eqs.~(\ref{growth1}) and
(\ref{evol_length}). The analysis of various quantities from numerical
simulations as metric distances or winding angles are all consistent with
the analytical theory. Interestingly, the first growth law (\ref{growth1})
does not contain the exponent $\nu$ and hence is superuniversal, being
the same for random and self-avoiding walks.  In the late stage dynamics
we have extended the results of ref.~\cite{wal13} to different models
and confirmed the scaling form of the longest relaxation time which
involves logarithmic corrections.

The emerging picture is that of a relatively quick loosening of the
polymer, which remains very close to the rod in the early stages of
the dynamics. This is followed by an intermediate regime where the
distance from the rod grows in a faster way and leads to the final
relaxation. Differently from the early and late time behaviors the
intermediate regime does not appear to display a clear cut scaling. 
This can be seen, for instance, in the behavior of
$R_{\rm e}^2$ depicted in Figure~\ref{fig:2}. During early dynamics, until a
characteristic time $\tau'$, $R_{\rm e}^2$ follows a power-law scaling. In the
late time dynamics $R_{\rm e}^2$ reaches a plateau. The intermediate time regime
links the two regimes showing no clear evidence of a power
law behavior. A typical snapshot of the intermediate
regime is shown in Figure~\ref{fig:1} (third configuration from the top). Its
characterization remains a challenge for future work.

The two phase model description of polymer dynamics has recently gained
some popularity: as examples we mention here the case of the translocation
of a polymer through a nanopore~\cite{sakaue07,rowg11,sakaue11}, the pulling
of a polymer by a constant force from one end~\cite{rowg12,sakaue12} and
the folding of a DNA hairpin~\cite{fred14}. In these problems
the polymer is subject to some local forcing and set into motion through
the propagation of tension along its backbone. As the tension does not
propagate instantaneously, the polymer is not set into motion at once. To
describe the motion it is usually assumed that one can divide the polymer
into different phases, which leads to some analytical predictions of the
exponents governing the dynamics~\cite{sakaue07,rowg11,sakaue11,fred14}. In
this paper a similar approach was adopted to study a complementary
case, namely that of the relaxation dynamics of an initially stretched
(helically wrapped) polymer. The excellent agreement between simulations
and model results shown in this paper, corroborates the validity of a
two phase model approach in the description of nonequilibrium
polymer dynamics.

\section*{Acknowledgement}
We thank G.T.~Barkema and A.~Revyakin for stimulating discussions. This
work is part of the research program of the Foundation for Fundamental
Research on Matter (FOM), which is financially supported by The
Netherlands Organisation for Scientific Research (NWO). J.-C.W. acknowledges
the support by the Laboratory of Excellence Initiative (Labex) NUMEV,
OD by the Scientific Council of the University of Montpellier 2 and
HPC@LR for offering computational resources.

\end{document}